\documentclass[
reprint,
amsmath,amssymb,
aps, prl,
]{revtex4-2}

\usepackage[title]{appendix}
\usepackage{graphicx}
\usepackage{dcolumn}
\usepackage{bm}
\usepackage{bbold}
\usepackage[hidelinks]{hyperref}
\usepackage[capitalise]{cleveref}
\usepackage{lipsum}
\usepackage[dvipsnames]{xcolor}
\usepackage{amsfonts}
\usepackage{yfonts}
\usepackage{booktabs}
\usepackage{tikz}
\usetikzlibrary{arrows.meta}

\usepackage[english]{babel}
\usepackage{amsthm}
\usepackage[scr=boondoxo,scrscaled=1.05]{mathalfa}
\theoremstyle{definition}

\newcommand{\rmd}{\mathrm{d}}
\newcommand{\Cov}{\operatorname{Cov}}
\newcommand{\diag}{\operatorname{diag}}
\newcommand{\SNR}{\operatorname{SNR}}

\DeclareFontFamily{U}{futm}{}
\DeclareFontShape{U}{futm}{m}{n}{
  <-> s * [.97534] fourier-bb 
  }{}
\DeclareMathAlphabet{\mathbbs}{U}{futm}{m}{n}

\begin{document}

\preprint{APS/123-QED}

\title{Spectral Duality and Thermodynamic Bounds on Finite-frequency Fluctuation Responses}
\author{Jiming Zheng} \email{jiming@unc.edu}
\author{Zhiyue Lu} \email{zhiyuelu@unc.edu}
\affiliation{Department of Chemistry, University of North Carolina-Chapel Hill, NC}

\date{\today}

\begin{abstract}
Fluctuation-response relations encode fundamental constraints on non-equilibrium systems. While time-domain static response is bounded by activity and entropy production, finite-frequency thermodynamic bounds for time-dependent perturbations remain largely unexplored. 
Here, we find a finite-frequency response-duality relation in non-equilibrium Markov jump processes. For state-current observables, the ratio between the spectral responses to kinetic-barrier and entropic-force perturbations is frequency-independent and gives the single-transition entropy production. The response-duality relation provides a method for measuring the single-transition entropy production from spectral response signals. Furthermore, we derive frequency-domain thermodynamic and kinetic inequalities for non-equilibrium systems with time-dependent perturbations around unperturbed steady states. We illustrate our response-duality relation on a quantum dot system. These finite-frequency response relations and inequalities provide a practical route for inferring dissipation from power-spectrum response measurements.

\end{abstract}

\maketitle

\emph{Introduction.}--- Understanding how systems respond to external perturbations is a fundamental challenge across physics, chemistry, and biology. The response, broadly defined, quantifies the deviation between the perturbed and unperturbed evolutions of a system, thereby encoding essential information about its internal dynamics. The most celebrated formulation is the linear response theory near equilibrium \cite{kubo1957statistical}, which establishes the connection between a near-equilibrium system's response to its internal fluctuations. Over the past two decades, trajectory-level formulations have extended response theory far from equilibrium, showing that non-equilibrium responses can be expressed through path-dependent correlation functions \cite{harada2005equality,speck2006restoring,seifert2010fluctuation,baiesi2009fluctuations,prost2009generalized}.

In recent years, significant progress has been made in extending the response theory to non-equilibrium processes. Specifically, the signal-to-noise ratio (SNR) of a linear response is bounded from above by kinetic quantities such as dynamical activity \cite{zheng2025universal,liu2025dynamical,kwon2025fluctuation}, as well as by thermodynamic quantities such as entropy production \cite{kwon2025fluctuation,aslyamov2025nonequilibrium,ptaszynski2024nonequilibrium,ptaszynski2025nonequilibrium} and cycle affinity \cite{owen2020universal}. These kinetic and thermodynamic bounds are known as Response-Kinetic Uncertainty Relations (R-KURs) and Response-Thermodynamic Uncertainty Relations (R-TURs), respectively. R-KURs and R-TURs have been established for various systems, including Markov jump processes, overdamped Langevin dynamics, and open quantum systems. However, most of these studies are limited to the time domain or static perturbations, with only a few addressing the spectral response relations \cite{aslyamov2026macroscopic} and time-dependent perturbations \cite{dechant2025finite}. In contrast, experiments are frequently performed to measure the frequency-domain response, including active matter \cite{mizuno2007nonequilibrium,netz2018fluctuation,turlier2016equilibrium,mizuno2008active}, molecular motors \cite{toyabe2010nonequilibrium,gieseler2021optical}, and cell mechanics \cite{nishizawa2017feedback,rigato2017high,gardel2008mechanical}. This raises a fundamental and practically relevant question: are there kinetic and thermodynamic bounds for finite-frequency responses under time-dependent perturbations? In particular, while kinetic bounds are already available for certain forms of spectral response \cite{dechant2025finite}, they do not discriminate between equilibrium and non-equilibrium conditions. A thermodynamic bound is indispensable for capturing the genuine non-equilibrium signature of a system and for enabling dissipation inference from spectral data.

In this Letter, we address this question by uncovering a finite-frequency thermodynamic structure of response around non-equilibrium steady states. We take the general finite-frequency FRI of Dechant \cite{dechant2025finite} as a starting point. Dechant's general inequality provides a kinetic constraint on spectral response, but by itself does not distinguish equilibrium from non-equilibrium steady states. Our central observation is that, for Markov jump processes and state-current observables, the responses to time-symmetric barrier perturbations and time-antisymmetric entropic force perturbations obey a universal duality relation. The ratio of these two responses at any frequency is fixed by the single-transition entropy production. This identity converts the general finite-frequency FRI into thermodynamic bounds involving the entropy production rate. As a result, the response to a kinetic barrier perturbation is constrained by dissipation at any frequency. Our results not only extend existing response relations to the frequency domain but also provide a practical pathway for estimating entropy production from experimentally accessible power spectra.

\emph{Setup.}--- We consider $n$-state Markov jump processes governed by the master equation
\begin{equation}
    \frac{\partial  p_i(t)}{\partial  t} = \sum_{i ( \neq j )} \left[ r_{ij} p_j(t) - r_{ji} p_i(t) \right],
    \label{eq: master equation}
\end{equation}
where $p_i(t)$ is the probability distribution in state $i$ at time $t$ and $r_{ij}$ is the transition rate from state $j$ to state $i$. Throughout this Letter,  we assume that the system is reversible, meaning that for every transition rate $r_{ij}$, the backward rate $r_{ji}$ is non-zero. We also assume the local detailed balance condition \cite{maes2021local} of the system, so that the entropy production is well-defined in the context of stochastic thermodynamics \cite{seifert2012stochastic,peliti2021stochastic}. The steady-state distribution $\{\pi_i\}$ is defined as the invariant distribution by
\begin{equation}
    0 = \frac{\partial \pi_i}{\partial t} = \sum_{i ( \neq j )} [r_{ij} \pi_j - r_{ji} \pi_i].
\end{equation}
The current on the edge $(i, j)$ is defined as $j_{ij} = r_{ij}\pi_j - r_{ji}\pi_i$; the all-direction transitions on this edge are characterized by the edge traffic $a_{ij} = r_{ij}\pi_j + r_{ji}\pi_i$. The dynamical activity of the system is defined as $a = \sum_{i < j} a_{ij}$ \cite{maes2020frenesy}. The dissipation of the system is characterized by the total entropy production rate (EPR) $\dot\sigma = \sum_{i < j} \dot\sigma_{ij}$, where $\dot\sigma_{ij} = j_{ij} \ln\frac{r_{ij}\pi_j}{r_{ji}\pi_i}$ is the entropy production rate on the edge $(i, j)$. For equilibrium steady states, all currents are zero, and thus the total EPR vanishes, $\dot\sigma = 0$. For non-equilibrium steady states, the EPR is strictly positive, $\dot\sigma > 0$.

We parameterize the transition rate $r_{ij}$ by the symmetric part $b_{ij} = b_{ji}$ and the antisymmetric part $f_{ij} = - f_{ji}$:
\begin{equation}
    r_{ij} = \exp \left[ b_{ij}(\zeta) + \frac{f_{ij}(\xi)}{2} \right],
    \label{eq: parameterization}
\end{equation}
where $\zeta$ and $\xi$ respectively control $\boldsymbol{b} = \{ b_{ij} \}$ and $\boldsymbol{f} = \{ f_{ij} \}$. For physical systems in contact with a thermal bath, the symmetric parameter $b_{ij}$ represents the energy barrier between states $i$ and $j$, which can be tuned by catalysts, magnetic fields \cite{hore2016radical,zadeh2022magnetic}, or nano-electronic techniques \cite{gustavsson2006counting,freitas2021stochastic}. The antisymmetric term $f_{ij}$ represents the change in entropy production in the thermal bath due to the transition $j \to i$, which can be affected by the change in the thermodynamic driving force and the energy landscape \cite{owen2020universal}. In this Letter, we consider time-dependent perturbations on the parameter $\lambda$ around unperturbed steady states: $\lambda \mapsto \lambda + \varepsilon \phi_\lambda(t)$, where $\phi_\lambda(t)$ is the temporal pattern of the perturbation with a small amplitude $\varepsilon \ll 1$. Throughout this letter, ``steady state" refers to the unperturbed reference process: a non-equilibrium steady state (NESS) of the original master equation. The time-dependent term $\varepsilon \phi_\lambda(t)$ is an infinitesimal probe defining the linear response around the unperturbed NESS. Without losing generality, here $\lambda$ can represent $b_{ij}$, $f_{ij}$, $\zeta$, or $\xi$.

To characterize the system's response, let us focus on state-current observables in the following form:
\begin{equation}
    Q(t) = \int_{\tau = 0}^{\tau = t} \sum_i g_i \rmd \tau_i(\tau) + \sum_{i \neq j} h_{ij} \rmd n_{ij}(\tau),
    \label{eq: observable}
\end{equation}
where $\rmd \tau_i(\tau)$ is the increment of dwelling time on the state $i$ accumulated in the infinitesimal time window $[\tau, \tau + \rmd \tau]$, and $\rmd n_{ij}(\tau)$ is the number of transitions in the same time window. Here, $h_{ij} = -h_{ji}$ are antisymmetric coefficients to capture edge-wise net currents. If $g_i = 0$ for all $i$, the observable $Q$ represents current type observables; while it represents state observables when $h_{ij} = 0$ for all $(i, j)$ pairs. In this Letter, we focus on its rate $\dot{Q} \equiv \rmd Q / \rmd t$. The fluctuation of $\dot{Q}(t)$ can be quantified by its spectral density function, defined as the Fourier transform of the autocorrelation function:
\begin{equation}
    \mathcal{S}(\omega) = \int_{-\infty}^{+\infty} e^{\mathrm{i}\omega t} \Cov(\dot{Q}(t), \dot{Q}(0)) \rmd t.
    \label{eq: spectral density}
\end{equation}

Typically, for general perturbations on a parameter $\lambda$, the steady-state linear response of the rate $\dot{Q}(t)$ can be captured by the response function:
\begin{equation}
    \delta \langle \dot{Q}(t) \rangle = \varepsilon \int_0^t R_\lambda(t-\tau) \phi_\lambda(\tau) \rmd \tau + o(\varepsilon),
    \label{eq: response function}
\end{equation}
where $\langle \cdots \rangle$ denotes the trajectory ensemble average. $R_\lambda(t - \tau) \equiv \frac{\delta\langle \dot{Q}(t)\rangle}{\delta\phi_\lambda(\tau)}$ describes the response on $\langle \dot{Q} \rangle$ at time $t$ to the perturbation on the parameter $\lambda$ at time $\tau$. The response function $R_\lambda(t)$ can also be described in the Fourier space
\begin{equation}
    \mathcal{R}_\lambda(\omega) = \int_0^{+\infty} e^{\mathrm{i} \omega \tilde{t}} R_\lambda(\tilde{t}) \rmd \tilde{t},
\end{equation}
where $\tilde{t} \equiv t - \tau$ and we use $R_\lambda(\tilde{t}) = 0$ for $\tilde{t} < 0$ due to causality.

\emph{Finite-frequency response-duality and fluctuation-response inequalities.}--- We first recall the finite-frequency FRI recently derived by Dechant \cite{dechant2025finite}, adapting it to the notation used here for a single measured observable and multiple perturbation parameters. This step provides the general spectral response inequality that will serve as the starting point for our thermodynamic construction. It is well known that a system's response is related to its fluctuations. In particular, it has been shown that the steady-state response function can be written as a correlation function \cite{speck2006restoring,seifert2010fluctuation,zheng2025unified,zheng2025nonlinear,dechant2025finite}
\begin{equation}
    R_\lambda(t - \tau) = \Cov(\dot{Q}(t), \dot{\Lambda}_\lambda(\tau)),
\end{equation}
where $\dot{\Lambda}_\lambda(\tau) = \left.\frac{\partial}{\partial \varepsilon} \frac{\delta \ln\mathcal{P}^\varepsilon[X_t]}{\delta \phi_\lambda(\tau)}\right|_{\varepsilon=0}$ can be specified as an observable, $\mathcal{P}[X_t]$ is the path probability, and $X_t$ is the stochastic path -- time series of the system's states and jump events in the time interval $[0, t]$. For Markov jump processes, the conjugate observable is $\rmd \Lambda_\lambda(t) = \sum_{i \neq j} \partial_\lambda \ln r_{ij} [\rmd n_{ij}(t) - r_{ij} \rmd \tau_j(t)]$, which is a linear combination of $\rmd n_{ij}$ and $\rmd \tau_j$. Notice that the steady-state condition makes sure the response function $R_\lambda$ only depends on the time difference $t - \tau$. For a vector of parameter $\boldsymbol{\lambda} = (\lambda_1, \cdots, \lambda_n)$ and conjugate observables $(\dot{\Lambda}_1, \cdots, \dot{\Lambda}_n)$, Dechant's finite-frequency FRI reads
\begin{equation}
    \boldsymbol{\mathcal{R}}(\omega) \boldsymbol{\mathcal{L}}^{-1}(\omega) \boldsymbol{\mathcal{R}}^\dagger(\omega) \le \mathcal{S}(\omega),
    \label{eq: block inequality}
\end{equation}
where ``$\dagger$'' denotes Hermitian transpose and $\boldsymbol{\mathcal{L}}(\omega)$ is the spectral density matrix of $\{\dot{\Lambda}_i\}$ with elements
\begin{equation}
    \mathcal{L}_{kl}(\omega) = \int_{-\infty}^{+\infty} e^{\mathrm{i}\omega t} \Cov(\dot{\Lambda}_k(t), \dot{\Lambda}_l(0)) \rmd t.
\end{equation}
Notice that both sides of \cref{eq: block inequality} are real numbers.

Now we consider the case where $\bm{\lambda}$ denotes $\{b_{ij}\}$ or $\{f_{ij}\}$. In this case, the matrix $\bm{\mathcal{L}}$ is diagonal and reads $\diag\{a_{ij}\}$ or $\diag\{a_{ij}/4\}$, respectively. Therefore, the finite-frequency FRIs for barrier and entropic perturbations are:
\begin{subequations}
\begin{align}
    \sum_{i < j} \frac{\left| \mathcal{R}_{b_{ij}}(\omega) \right|^2}{a_{ij}} &\le \mathcal{S}(\omega), \label{eq: FRIs 1a} \\
    \sum_{i < j} \frac{4 \left| \mathcal{R}_{f_{ij}}(\omega) \right|^2}{a_{ij}} &\le \mathcal{S}(\omega),
    \label{eq: FRIs 1b}
\end{align}
\end{subequations}
where $\left| \mathcal{R}_{\lambda}(\omega) \right|^2 = \mathcal{R}_{\lambda}(\omega) \mathcal{R}^\dagger_{\lambda}(\omega)$. \cref{eq: FRIs 1a,eq: FRIs 1b} are the edge-resolved, barrier/force specialization of Dechant's finite-frequency FRI \cite{dechant2025finite}; we reproduce them here only as the kinetic starting point. The new, thermodynamic content of this Letter begins with the response-duality relation \cref{eq: response ratio,eq: response ratio entropy}.

The above FRIs generally hold for arbitrary trajectory observables and do not involve non-equilibrium information of the system. The key additional ingredient is the response dual-relation below. For state-current observables of the form \cref{eq: observable}, we prove in the Supplemental Material \cite{supp} that the finite-frequency responses to the barrier parameter $b_{ij}$ and the entropic parameter $f_{ij}$ are related through
\begin{equation}
    \frac{\mathcal{R}_{b_{ij}}(\omega)}{\mathcal{R}_{f_{ij}}(\omega)} = \frac{2 j_{ij}}{a_{ij}}.
    \label{eq: response ratio}
\end{equation}
\cref{eq: response ratio} holds for state-current observables in steady states at any frequency. The previously discovered time-domain relation in \cite{kwon2025fluctuation} is the zero-frequency limit of \cref{eq: response ratio}.
The response to barrier perturbation $\mathcal{R}_{b_{ij}}(\omega)$ vanishes at equilibrium and becomes nonzero only in the presence of net current on the edge.

Let us define the steady-state thermodynamic force on the edge $j \to i$ by
\begin{equation}
    A_{ij} \equiv \ln\frac{r_{ij}\pi_j}{r_{ji}\pi_i}.
\end{equation}
It is anti-symmetric, i.e., $A_{ij} = -A_{ji}$. Under the local detailed balance condition \cite{maes2021local}, $A_{ij}$ represents the entropy change caused by the single transition event $j \to i$. Its average over transition numbers gives the entropy production on the edge $(i, j)$: $\sigma_{ij} = \langle A_{ij}n_{ij} + A_{ji}n_{ji} \rangle$. In this case, \cref{eq: response ratio} can be rewritten in terms of $A_{ij}$ as
\begin{equation}
    \frac{\mathcal{R}_{b_{ij}}(\omega)}{\mathcal{R}_{f_{ij}}(\omega)} = 2 \tanh \frac{A_{ij}}{2}.
    \label{eq: response ratio entropy}
\end{equation}
Remarkably, the ratio is independent of frequency $\omega$, although each response function has nontrivial frequency dependence. Thus, the spectral response ratio directly measures the edge thermodynamic force $A_{ij}$, or equivalently, the stochastic entropy increment associated with a single transition. Multiplying this local entropy increment by the corresponding steady current gives the edge contribution to the entropy production rate. Notice that the ratio in \cref{eq: response ratio entropy} is obtained from two directly measurable response spectra and requires no prior knowledge of the transition rates. It thus extracts an {\it a priori} unknown thermodynamic entropy production from measurable response data, rather than presuming the full kinetic model.

\cref{eq: response ratio entropy} is our first main result. It allows us to infer the single-transition entropy production from response measurements. Near equilibrium, the fluctuation-dissipation theorem states that only the imaginary part of a response function is related to dissipation \cite{kubo1957statistical}. In contrast, for non-equilibrium steady states, \cref{eq: response ratio entropy} implies that both the real and imaginary parts are related to the thermodynamic irreversibility.

\cref{eq: response ratio} introduces thermodynamic irreversibility into the finite-frequency response theory. In particular, it allows the kinetic FRIs in \cref{eq: FRIs 1a,eq: FRIs 1b} to be transformed into bounds involving net currents:
\begin{subequations}
\begin{align}
    \sum_{i < j} \frac{a_{ij} \left| \mathcal{R}_{b_{ij}}(\omega) \right|^2}{{j_{ij}}^2} &\le \mathcal{S}(\omega),
    \label{eq: FRIs 2a} \\
    \sum_{i < j} \frac{4 {j_{ij}}^2 \left| \mathcal{R}_{f_{ij}}(\omega) \right|^2}{{a_{ij}}^3} &\le \mathcal{S}(\omega).
    \label{eq: FRIs 2b}
\end{align}
\end{subequations}

\emph{Kinetic and thermodynamic bounds.}--- The above FRIs further lead to the frequency-domain R-KURs and R-TURs. Defining the frequency-domain signal-to-noise ratio as
\begin{equation}
    \SNR_\lambda(\omega) \equiv \frac{|\mathcal{R}_\lambda(\omega)|^2}{\mathcal{S}(\omega)}.
    \label{eq: SNR1}
\end{equation}
Let us consider perturbations in parameters $\zeta$ and $\xi$, which globally control the kinetic barrier $\{ b_{ij} \}$ and the entropic force $\{ f_{ij} \}$, respectively. Applying the chain rule and Cauchy-Schwarz inequality $\sum_{i<j} (x_{ij}/y_{ij})^2 \ge (\sum_{i<j} x_{ij})^2 / (\sum_{i<j} {y_{ij}}^2)$ to \cref{eq: FRIs 1a,eq: FRIs 1b}, we obtain the following inequalities:
\begin{subequations}
\begin{align}
    \mathcal{S}(\omega) \ge \frac{\left| \mathcal{R}_\zeta(\omega) \right|^2}{\displaystyle{\sum_{i<j} ( \partial_\zeta b_{ij} )^2 a_{ij}}} \ge \frac{\left| \mathcal{R}_\zeta(\omega) \right|^2}{\displaystyle{\max_{(i, j)} \left\{( \partial_\zeta b_{ij} )^2 \right\} \cdot a}}, \label{eq: Cauchy zeta} \\
    \mathcal{S}(\omega) \ge \frac{4 \left| \mathcal{R}_\xi(\omega) \right|^2}{\displaystyle{\sum_{i<j} ( \partial_\xi f_{ij} )^2 a_{ij}}} \ge \frac{4 \left| \mathcal{R}_\xi(\omega) \right|^2}{\displaystyle{\max_{(i, j)} \left\{( \partial_\xi f_{ij} )^2 \right\} \cdot a}}, \label{eq: Cauchy xi}
\end{align}
\end{subequations}
where we choose $x_{ij} = (\partial_\zeta b_{ij}) \sqrt{|\mathcal{R}_{b_{ij}}(\omega)|^2}$ and $y_{ij} = (\partial_\zeta b_{ij}) \sqrt{a_{ij}}$ in \cref{eq: Cauchy zeta}, and $x_{ij} = (\partial_\xi f_{ij}) \sqrt{|\mathcal{R}_{f_{ij}}(\omega)|^2}$ and $y_{ij} = (\partial_\xi f_{ij}) \sqrt{a_{ij}}/2$ in \cref{eq: Cauchy xi}.
The straightforward rearrangements of \cref{eq: Cauchy zeta,eq: Cauchy xi} lead to finite-frequency R-KURs for general observables:
\begin{subequations}
\begin{align}
    \SNR_\zeta(\omega) &\le \max_{(i, j)} \left\{ (\partial_\zeta b_{ij})^2 \right\} \cdot a, \label{eq: R-KUR a} \\
    \SNR_\xi(\omega) &\le \max_{(i, j)} \left\{ (\partial_\xi f_{ij})^2 \right\} \cdot \frac{a}{4}. \label{eq: R-KUR b}
\end{align}
\end{subequations}
Finite-frequency R-KURs state that the response of any trajectory observable on kinetic barrier and entropic force perturbations is bounded from above by the system's dynamical activity. More active systems, i.e., systems with a larger number of transitions per unit time, can allow for larger responses in the frequency domain.

The thermodynamic bounds, finite-frequency R-TURs, can be obtained by applying the chain rule and Cauchy-Schwarz inequality to \cref{eq: FRIs 2a}. It yields an inequality for $\mathcal{R}_\zeta(\omega)$ on state-current type observables:
\begin{align}
    \mathcal{S}(\omega) &\ge \frac{\left| \mathcal{R}_\zeta(\omega) \right|^2}{\displaystyle{\sum_{i<j} (\partial_\zeta b_{ij})^2 {j_{ij}}^2/a_{ij}}} \nonumber \\
    &\ge \frac{2 \left| \mathcal{R}_\zeta(\omega) \right|^2}{\displaystyle{\max_{(i, j)} \left\{ (\partial_\zeta b_{ij})^2 \right\} \cdot \dot\sigma^{\text{pseudo}}}}, \label{eq: Cauchy pseudo}
\end{align}
where $\dot\sigma^{\text{pseudo}} \equiv \sum_{i<j} 2 {j_{ij}}^2/a_{ij}$ is the pseudo-EPR \cite{shiraishi2021optimal}. The rearrangement leads to finite-frequency R-TURs:
\begin{align}
    \SNR_\zeta(\omega) &\le \max_{(i, j)} \left\{ (\partial_\zeta b_{ij})^2 \right\} \cdot \frac{\dot\sigma^{\text{pseudo}}}{2} \nonumber \\
    &\le \max_{(i, j)} \left\{ (\partial_\zeta b_{ij})^2 \right\} \cdot \frac{\dot\sigma}{2}. \label{eq: R-TUR}
\end{align}
The second inequality follows from $\dot\sigma^{\text{pseudo}} \le \dot\sigma$, which is given by the log-mean inequality $2(x_{ij}-y_{ij})^2/(x_{ij}+y_{ij}) \le (x_{ij} - y_{ij})(\ln x_{ij} - \ln y_{ij})$ with $x_{ij} = r_{ij}\pi_j$ and $y_{ij} = r_{ji}\pi_i$. The finite-frequency R-TUR \cref{eq: R-TUR} is our second main result. \cref{eq: R-TUR} only holds for state-current type observables since the derivation relies on the response relation \cref{eq: response ratio}, which is unique to observables of the form  \cref{eq: observable}. The thermodynamic bound states that the response to barrier perturbations is bounded from above by the system's dissipation, which captures the system's non-equilibrium nature. Therefore, the finite-frequency response to barrier perturbations is generally related to how far the system is from equilibrium. We also derive the frequency thermodynamic inequalities for overdamped Langevin systems in the End Matter.

\begin{figure}[htbp]
  \centering
  \includegraphics[width=0.8\linewidth]{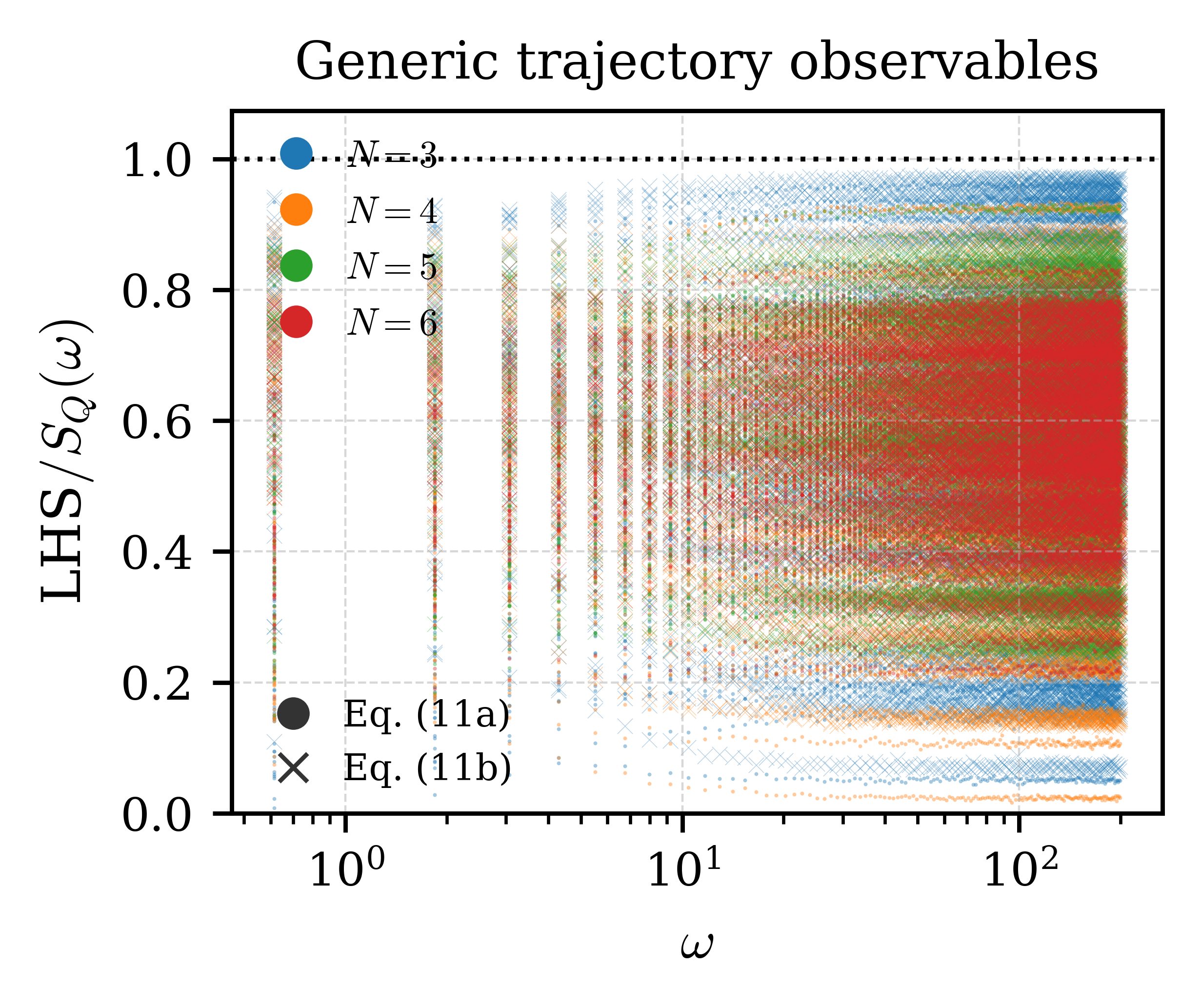}
  \vspace{-0.5cm}
  \includegraphics[width=0.8\linewidth]{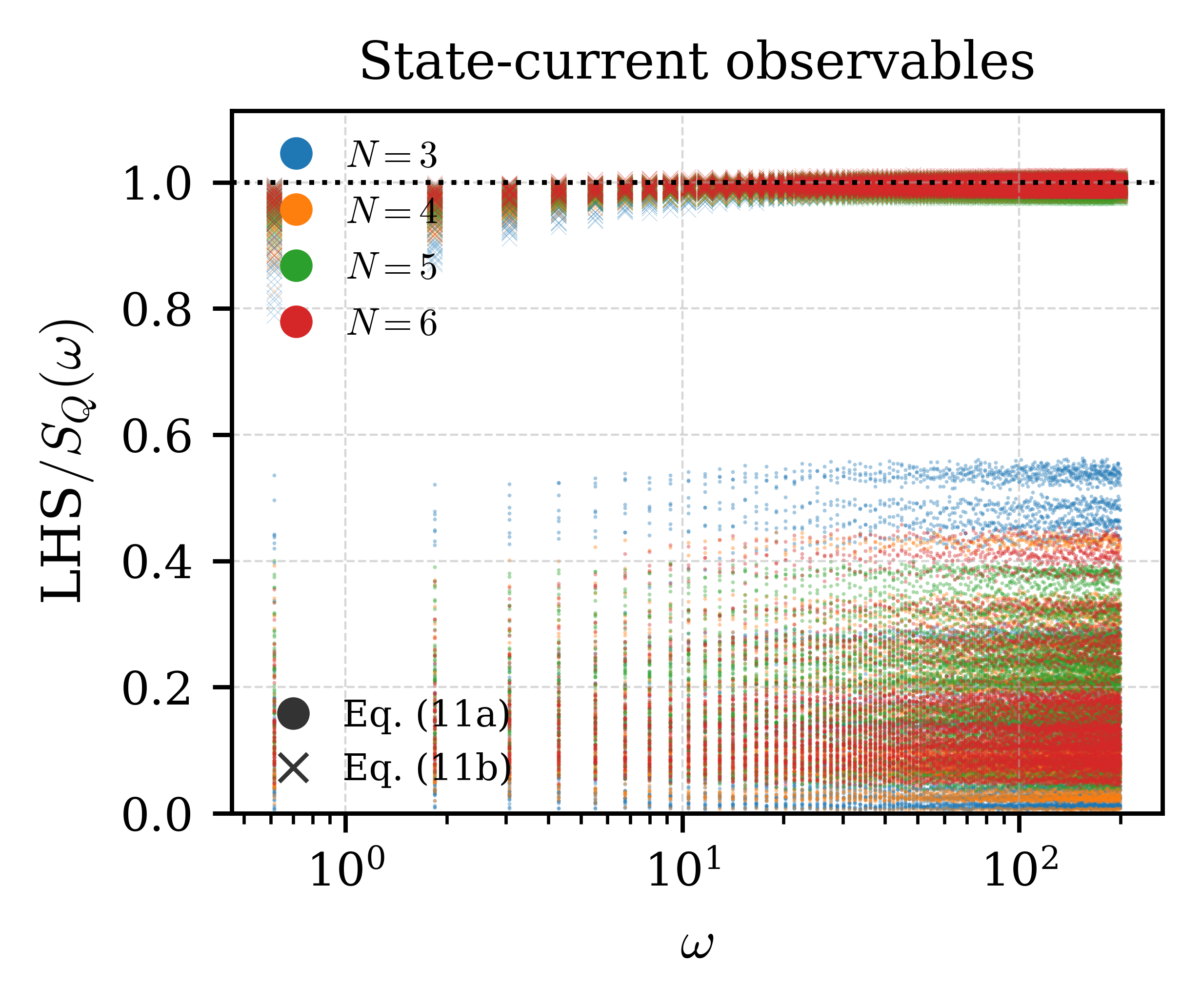}
  \caption{Numerical verification of \cref{eq: FRIs 1a,eq: FRIs 1b}. Normalized bounds $\mathrm{LHS}/S_Q(\omega)$ for randomly sampled Markov jump processes with different state-space sizes, transition rates, observables, and frequencies. The upper panel corresponds to generic trajectory observables, and the lower panel corresponds to state-current observables. Colors denote network size, and marker shapes distinguish the two inequalities.}
  \label{fig:example}
\end{figure}

\emph{Tightness and saturation conditions.}--- Our two main relations behave very differently with respect to tightness. The response-duality relation \cref{eq: response ratio entropy} is an \emph{equality} and is therefore exactly tight by construction. As shown later in numerical demonstrations, its convergence is limited only by the statistical accuracy of the measured responses. The finite-frequency R-TUR \cref{eq: R-TUR}, by contrast, chains three inequalities: (i) \cref{eq: FRIs 1b} or equivalently \cref{eq: FRIs 2a}, (ii) $(\partial_\zeta b_{ij})^2 \le \max_{(i, j)}(\partial_\zeta b_{ij})^2$, and (iii) $\dot{\sigma}^{\text{pseudo}} \le \dot{\sigma}$. As shown in \cref{fig:example}, step (i) \cref{eq: FRIs 1b} saturates for state-current observables. Despite the numerical evidence in \cref{fig:example}, its saturation has recently been proved in \cite{aslyamov2026dynamical,kwon2026nonequilibrium}. Step (ii) is saturated when the global perturbation $\zeta$ couples uniformly to all barriers $\{b_{ij}\}$. Step (iii) is saturated in the equilibrium limit, i.e., $\dot{\sigma^{\text{pseudo}}} \approx \dot{\sigma} \to 0$. Therefore, the finite-frequency R-TUR is tight when the effect of global perturbation on $\zeta$ is approximately uniform, and the steady state is close to equilibrium.

\vspace{-0.5cm}

\emph{Application: quantum dot.}--- We now illustrate the finite-frequency response-duality relation using a spinful quantum dot coupled to two electronic reservoirs. Quantum dots provide a natural mesoscopic platform for stochastic thermodynamics: their charge states and tunneling events can be monitored in real time by nearby charge sensors, and full counting statistics of electron transport have been measured experimentally \cite{lu2003real,gustavsson2006counting,fujisawa2004electron}. Moreover, tunnel-barrier gates, source-drain bias, plunger gates, and magnetic fields provide independent experimental handles on the kinetic couplings, thermodynamic driving, and spin-dependent level structure \cite{simmons2009charge,hanson2007spins,amasha2008spin,hanson2005single}. These features make quantum dots a suitable platform for testing response relations between kinetic and entropic perturbations.

\begin{figure}[htbp]
    \centering
    \includegraphics[width=0.99\linewidth]{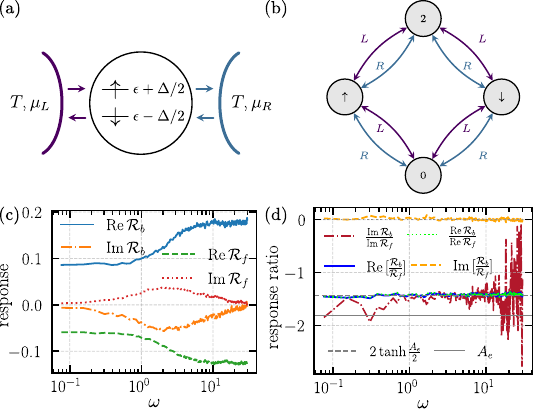}
    \caption{Finite-frequency response-duality in a channel-resolved spinful quantum dot.
    (a) Schematic of a single-level quantum dot coupled to two electronic reservoirs $L$ and $R$. The dot has four states $0,\uparrow,\downarrow,2$, with energies $E_0 = 0$, $E_\uparrow = \epsilon-\Delta/2$, $E_\downarrow = \epsilon+\Delta/2$, and $E_2 = 2\epsilon + U$.
    (b) Equivalent channel-resolved Markov network. Tunneling through $L$ and $R$ defines distinct jump channels; channels with the same initial and final dot states are summed in the four-state master equation, $r_{ij} \equiv \sum_{\alpha=L,R}r_{ij}^{\alpha}$. The channel rates are $r_{\sigma,0}^{\alpha} = \Gamma_{\alpha\sigma}^{(0)} f_\alpha (E_\sigma-E_0)$, $r_{0,\sigma}^{\alpha} = \Gamma_{\alpha\sigma}^{(0)} [1-f_\alpha(E_\sigma-E_0)]$, $r_{2,\sigma}^{\alpha} = \Gamma_{\alpha\bar\sigma}^{(1)} f_\alpha (E_2-E_\sigma)$, and $r_{\sigma,2}^{\alpha} = \Gamma_{\alpha\bar\sigma}^{(1)} [1-f_\alpha(E_2-E_\sigma)]$, where $f_\alpha(E) = 1/[1+\exp{\beta(E-\mu_\alpha)}]$. Parameters: $\beta=1$, $\epsilon=0$, $U=2.0$, $\Delta_Z=0.6$, $\mu_L=2.5$, $\mu_R=-1.5$, and $(\Gamma_{L\uparrow}^{(0)}, \Gamma_{L\downarrow}^{(0)}, \Gamma_{R\uparrow}^{(0)}, \Gamma_{R\downarrow}^{(0)}, \Gamma_{L\uparrow}^{(1)}, \Gamma_{L\downarrow}^{(1)}, \Gamma_{R\uparrow}^{(1)},\Gamma_{R\downarrow}^{(1)}) = (1.0, 0.7, 0.6, 1.1, 0.8, 1.2, 1.0, 0.75)$. For each channel edge $e$, the positive direction is chosen as electron tunneling from the reservoir into the dot, and $r_e^\pm=\exp(b_e\pm f_e/2)$.
    (c) Real and imaginary parts of the response spectra $\mathcal{R}_b(\omega)$ and $\mathcal{R}_f(\omega)$ for the perturbed channel $e$ and observable $J_R$.
    (d) Verification of \cref{eq: response ratio entropy}: the ratios $\operatorname{Re} \mathcal{R}_b / \operatorname{Re} \mathcal{R}_f$, $\operatorname{Im} \mathcal{R}_b / \operatorname{Im} \mathcal{R}_f$, and $\operatorname{Re}[\mathcal{R}_b/\mathcal{R}_f]$ collapse onto the frequency-independent value $2\tanh(A_e/2)$.}
    \label{fig:quantum_dot}
\end{figure}

We use a channel-resolved Markov description, where tunneling through the left and right reservoirs is treated as distinct jump channels. Channels connecting the same pair of dot states are summed as transition rates in the master equation, while trajectory observables and conjugate response observables retain the channel label. The multi-channel description is equivalent to the multiple reservoir cases in stochastic thermodynamics \cite{kung2012irreversibility,peliti2021stochastic}. We take the measured observable to be the net particle current into the right reservoir, $J_R$. The perturbed edge $e$ is the transition between empty state $(0)$ and down state $(\downarrow)$ coupled with the right reservoir $(R)$. As shown in \cref{fig:quantum_dot}(c), the barrier and force responses, $\mathcal{R}_b(\omega)$ and $\mathcal{R}_f(\omega)$, have nontrivial frequency dependence. Nevertheless, their ratio is fixed by $2\tanh(A_e/2)$ over the full frequency range, as shown in \cref{fig:quantum_dot}(d). The ratio of real parts, $\operatorname{Re} \mathcal{R}_b / \operatorname{Re} \mathcal{R}_f$ converges well around $2\tanh(A_e/2)$, while the ratio of imaginary parts, $\operatorname{Im} \mathcal{R}_b / \operatorname{Im} \mathcal{R}_f$, fluctuates a lot due to the small norm of $\operatorname{Im}\mathcal{R}_f$. Therefore, \cref{eq: response ratio entropy} offers an experimental method to measure the single-transition entropy production through spectral response measurements.

\emph{Concluding remarks.}--- In this Letter, we uncovered a finite-frequency response-duality relation in non-equilibrium Markov jump processes --- the ratio between barrier and entropic force responses for state-current observables is related to the single-transition entropy production. It provides an experimental method to measure the single-transition entropy production from spectral response data. It further implies finite-frequency R-TURs on barrier perturbations for state-current observables, and provides a method to infer the EPR from spectral response measurements.

Our spectral bounds for Markov jump systems extend time-domain inequalities in a nontrivial way. The finite-frequency R-TURs arise from the delta-correlated nature of the Markovian noise and the response-duality relation between different types of perturbations. The first ingredient renders the spectral density $\bm{\mathcal{L}}$ frequency-independent and diagonal \cite{dechant2025finite}, while the second one introduces thermodynamic irreversibility into the frequency regime. Several interesting open questions emerge from the analysis. Extending these ideas to nonlinear response and non-Markovian systems would be interesting and practically valuable. It is also interesting to generalize our theory to macroscopic systems and open quantum systems, where time-domain or static response theories have recently been uncovered \cite{zheng2025nonequilibrium,aslyamov2026macroscopic,kwon2025fluctuation,liu2025response}.

\emph{Note added.}--- During the peer review process of this manuscript, two related preprints appeared, where the equality conditions of \cref{eq: FRIs 1b,eq: FRIs 2a} of the present work were discussed \cite{aslyamov2026dynamical,kwon2026nonequilibrium}.

\emph{Acknowledgments.}--- J. Z. thanks Kecai Xuan for helpful discussions on numerical simulations. This work is supported by the U.S. National Science Foundation under Grant No. DMR-2145256 and Alfred P. Sloan Foundation Matter-to-Life Theory Award under Grant No. G-2025-25194.

\emph{Data availability.}--- The data that support the findings of this article are generated by numerical simulation codes that are openly available
at \cite{data}.

\appendix
\section{End Matter}

\setcounter{equation}{0}
\renewcommand{\theequation}{A\arabic{equation}}

\emph{Experimental measurement of the spectral density.}--- We now discuss how the spectral density $\mathcal{S}(\omega)$ can be obtained from experimental trajectory data. The quantity $\mathcal{S}(\omega)$ is the power spectral density of the fluctuating rate $\dot Q(t)$ in the unperturbed steady state defined as \cref{eq: spectral density}. Its measurement does not require prior knowledge of the transition-rate matrix. It only requires repeated measurements of the stochastic trajectories from which the observable $Q(t)$ can be constructed. Such trajectory-level measurements are routinely performed in single-molecule experiments, colloidal systems, optical-trap experiments, and feedback-controlled Langevin systems \cite{jun2012virtual,gavrilov2017feedback,kumar2018nanoscale,albay2018optical}.

For a Markov jump process, an experimental trajectory consists of a sequence of states and jump times. Given such a trajectory, the state-current observable \cref{eq: observable} can be constructed. In practice, one records $N_{\mathrm{traj}}$ independent steady-state trajectories of duration $T_{\mathrm{obs}} = N_t\Delta t$ and bins the observable into increments
\begin{equation}
    q_m^{(\alpha)} \equiv \frac{Q^{(\alpha)}((m+1)\Delta t) - Q^{(\alpha)}(m\Delta t)}{\Delta t},
\end{equation}
where $\alpha$ labels the trajectory and $ m = 0, \cdots, N_t-1$. The steady-state mean is estimated by
\begin{equation}
  \bar{q} = \frac{1}{N_{\mathrm{traj}}N_t} \sum_{\alpha=1}^{N_{\mathrm{traj}}} \sum_{m=0}^{N_t-1} q_m^{(\alpha)},
\end{equation}
and the fluctuation is $\delta q_m^{(\alpha)} = q_m^{(\alpha)} - \bar{q}$. The periodogram average gives a direct estimator of the spectral density:
\begin{equation}
  \mathcal{S}(\omega_j) = \frac{\Delta t}{N_t N_{\mathrm{traj}}} \sum_{\alpha=1}^{N_{\mathrm{traj}}} \left| \sum_{m=0}^{N_t-1} \delta q_m^{(\alpha)} e^{\mathrm{i}\omega_j m\Delta t} \right|^2 ,
  \label{SMeq: spectral_density_estimator}
\end{equation}
where $\omega_j=2\pi j/(N_t\Delta t)$ are the discrete Fourier frequencies. Equivalently, one may first estimate the steady-state autocovariance of $\dot{Q}$ and then Fourier transform it. Eq. \eqref{SMeq: spectral_density_estimator} is simply the trajectory-level implementation of the defining relation for $\mathcal{S}(\omega)$.

\emph{Frequency integrated inequalities.}--- The signal-to-noise ratio in the main text defines noise by the spectral density of the observable $\mathcal{S}(\omega)$. Alternatively, one can define noise by the fluctuations in the time-domain, $\operatorname{Var}[\dot{Q}]$, which is easier to obtain than $\mathcal{S}(\omega)$ experimentally. To connect the time-domain fluctuations to the finite-frequency inequality, below we consider the alternative signal-to-noise ratio involving $\operatorname{Var}[\dot{Q}]$:
\begin{equation}
    \operatorname{SNR}'_\lambda(\omega) \equiv \frac{|\mathcal{R}_\lambda(\omega)|^2}{\operatorname{Var}[\dot{Q}]},
\end{equation}
where the ``$\prime$'' distinguishes the above SNR from the frequency-domain SNRs in the main text.

Integrating $\mathcal{S}(\omega)$ over the frequency domain gives the time-domain fluctuations:
\begin{equation}
    \frac{1}{\pi} \int_0^\infty \mathcal{S}(\omega) \mathrm{d} \omega = \operatorname{Var}[\dot{Q}].
\end{equation}
Therefore, integrating both sides of Eqs. (17a), (17b), and (19), and rearrangements give the following kinetic and thermodynamic inequalities for integrated $\operatorname{SNR}'_\lambda(\omega)$:
\begin{subequations}
\begin{align}
    \frac{1}{\pi} \int_0^\infty \operatorname{SNR}'_\zeta(\omega) \mathrm{d}\omega &\le \max_{(i, j)} \left\{ (\partial_\zeta b_{ij})^2 \right\} \cdot a, \\
    \frac{1}{\pi} \int_0^\infty \operatorname{SNR}'_\xi(\omega) \mathrm{d}\omega &\le \max_{(i, j)} \left\{ (\partial_\xi f_{ij})^2 \right\} \cdot \frac{a}{4}, \\
    \frac{1}{\pi} \int_0^\infty \operatorname{SNR}'_\zeta(\omega) \mathrm{d}\omega &\le \max_{(i, j)} \left\{ (\partial_\zeta b_{ij})^2 \right\} \cdot \frac{\dot{\sigma}}{2}.
\end{align}
\end{subequations}
These integrated inequalities suggest a constraint on the total spectral weight of response precision. When the integrated bound is close to saturation, enhancing the response precision in one frequency window must be compensated for by reductions elsewhere.

\emph{Overdamped Langevin systems.}--- In parallel to the Markov jump process, similar thermodynamic and kinetic bounds for overdamped Langevin systems can be extended from the time domain \cite{chun2026fluctuation} to the frequency domain. Since the recent study \cite{chun2026fluctuation} proves thermodynamic inequalities for time-dependent perturbations, the finite-frequency counterparts are direct results of its Fourier transform. Consider the overdamped Langevin equation
\begin{equation}
    \dot{x}(t) = \mu(x_t) F(x_t) + \sqrt{2\mu(x_t)T(x_t)} \star \xi(t),
\end{equation}
where $\mu(x)$ is the mobility and ``$\star$'' stands for anti-Ito product, which ensures the thermodynamic consistency for systems with multiplicative noise \cite{lau2007state}. We consider the following state-current observable:
\begin{equation}
    Q(x, t) = \int_{\tau=0}^{\tau=t} \left[u(x_\tau) + \dot{x}_\tau \circ w(x_\tau)\right] \rmd \tau,
\end{equation}
where $u(x)$ and $w(x)$ are arbitrary functions, and $\circ$ denotes the Stratonovich product. The steady-state average of its rate is given by $\langle \dot{Q} \rangle = \int [u(x)\pi(x) + w(x)j_\text{ss}(x)] \rmd x$, where $\pi(x)$ is the steady-state distribution and $j_\text{ss}(x) \equiv \mu(x)[F(x)\pi(x) - T(x)\partial_x \pi(x)]$ is the steady-state probability current.

We consider the time-dependent perturbation on a local parameter $\lambda(x) \mapsto \lambda(x) + \varepsilon\phi_\lambda(x, t)$ around an unperturbed steady state. This local parameter could represent $\mu(x)$, $F(x)$, and $T(x)$. Similarly to the discrete case, we also consider a global parameter $\gamma$ that controls the local parameter $\lambda(x, \gamma)$ globally. The steady-state response of the observable $Q$ is characterized by
\begin{equation}
    \delta\langle \dot{Q}(\tau) \rangle = \varepsilon \int_0^\tau R_\lambda(x, \tau-t) \phi_\lambda(x, t) \rmd t + o(\varepsilon),
\end{equation}
where $R_\lambda(x, \tau-t) \equiv \frac{\delta\langle \dot{Q}(\tau) \rangle}{\delta \phi_\lambda(x, t)}$ is the steady-state response function. It can also be written as a correlation function $R_\lambda(x, \tau - t) = \langle \dot{Q}(x, \tau) \dot{\Lambda}_\lambda(x, t) \rangle$. Its Fourier transform $\mathcal{R}_\lambda(x, \omega)$ satisfies the finite-frequency inequalities (see \cite{supp} for detailed derivations):
\begin{subequations}
\begin{align}
    \int \frac{\left|  \mathcal{R}_\lambda(x, \omega) \right|^2}{\mathcal{L}_\lambda(x)} \rmd x &\le \mathcal{S}(\omega), \label{eq: Langevin FRIs} \\
    \frac{\int \left|  \mathcal{R}_\gamma(x, \omega) \right|^2 \rmd x}{\mathcal{S}(\omega)} &\le \sup_{x}\{ (\partial_\gamma \lambda)^2 \} \cdot \int \mathcal{L}_\lambda(x) \rmd x, \label{eq: SRN bound langevin}
\end{align}
\end{subequations}
where $\mathcal{L}_\lambda \equiv \int_{-\infty}^{+\infty} e^{\mathrm{i}\omega t}\operatorname{Cov}(\dot{\Lambda}(x, t), \dot{\Lambda}(x', 0)) \rmd t$ is the spectral density of $\dot{\Lambda}_\lambda(x, t)$. \cref{eq: SRN bound langevin} follows from \cref{eq: Langevin FRIs} by applying the Cauchy-Schwarz inequality. Similarly to the discrete case, $\mathcal{L}_\lambda$ is independent of $\omega$ and contains the space delta function $\delta(x - x')$ \cite{supp}. The explicit expressions of $\mathcal{L}_\lambda$ for some specific $\lambda$s are listed in \cref{tab: parameter-FI}. Note that when $\lambda = \ln\mu$, $\mathcal{L}_{\ln\mu} = \int \frac{{j_\text{ss}}^2}{2\mu T \pi} \rmd x \equiv \dot{\sigma}$ is the steady state EPR for the Langevin system. So \cref{eq: SRN bound langevin} extends the thermodynamic bound on the response of the Langevin system \cite{chun2026fluctuation} to the finite-frequency regime.

\begin{table}[htbp]
    \centering
    \begin{tabular}{cc}
        \toprule
        $\lambda(x)$ & $\mathcal{L}_\lambda(x)$ \\
        \midrule
        $\ln\mu$ & ${j_\text{ss}}^2 / (2\pi \mu T)$ \\
        $F$ & $\pi\mu / (2T)$ \\
        $T$ & $\pi\mu(\partial_x \ln\pi)^2 / (2T)$ \\
        \bottomrule
    \end{tabular}
    \caption{Perturbed parameter $\lambda(x)$ and its corresponding spectral density $\mathcal{L}_\lambda(x)$}
    \label{tab: parameter-FI}
\end{table}

\bibliography{manuscript}

\end{document}


\preprint{APS/123-QED}

\title{Supplementary Material for ``Spectral Duality and Thermodynamic Bounds on Finite-frequency Fluctuation Responses''}

\author{Jiming Zheng}
 \email{jiming@unc.edu}
\author{Zhiyue Lu}%
 \email{zhiyuelu@unc.edu}
\affiliation{%
Department of Chemistry, University of North Carolina-Chapel Hill, NC}

\date{\today}

\maketitle

In this Supplementary Material, we provide complementary derivations of the results for discrete and continuous systems. We also provide some details of the simulations in the main text.

\tableofcontents

\newpage

\section{Derivations for discrete systems} \label{SIsec: inequality master}

\subsection{\texorpdfstring{Derivation of $\boldsymbol{\mathcal{L}}(\omega)$}{Derivation of L(w)}} \label{SIsec: Hermitian}

We derive the explicit expression of $\boldsymbol{\mathcal{L}}(\omega)$ for $\{ b_{ij} \}$ and $\{ f_{ij} \}$ from the stochastic calculus of the master equation. For completeness, we briefly review the stochastic calculus framework here. Please see \cite{stutzer2025stochastic} for detailed descriptions. Similar to the Fokker-Planck and Langevin descriptions, the master equation also has a corresponding stochastic differential equation (SDE):
\begin{equation}
    \mathrm{d} n_{ij}(\tau) = r_{ij} \mathrm{d} \tau_j(\tau) + \mathrm{d} \varepsilon_{ij}(\tau),
    \label{SIeq: master SDE}
\end{equation}
where $\mathrm{d} n_{ij}(\tau) = 0$ or $1$ is the number of transitions, $\mathrm{d} \tau_j(\tau) = 0$ or $\mathrm{d} \tau$ is the dwelling time in the state $j$, and $\mathrm{d} \varepsilon_{ij}(\tau)$ is the noise of the jump process in the infinitesimal time window $\mathrm{d} \tau$. Different from the Gaussian white noise in the Langevin equation, the noise $\mathrm{d} \varepsilon_{ij}(\tau)$ is a Poisson noise defined as
\begin{equation}
    \mathrm{d} \varepsilon_{ij}(\tau) \equiv \mathrm{d} n_{ij}(\tau) - r_{ij}\mathrm{d} \tau_j(\tau),
\end{equation}
with first and second order moments
\begin{subequations}
\begin{align}
    \langle \mathrm{d} \varepsilon_{ij}(\tau) \rangle &= 0, \label{SIeq: 1st moment} \\
    \langle \mathrm{d} \varepsilon_{ij}(\tau_1) \mathrm{d} \varepsilon_{kl}(\tau_2) \rangle &= \delta_{ik} \delta_{jl} \delta(\tau_1-\tau_2) r_{ij} p_j \mathrm{d} \tau_1 \mathrm{d} \tau_2. \label{SIeq: 2nd moment}
\end{align}
\end{subequations}

A stochastic path $X_t$ is fully determined by the sequence of states $\{ x_i \}$ and transition times $\{ t_i \}$. In the SDE language, the path probability generated by the master equation can be written as
\begin{equation}
    \mathcal{P}[X_t] = p_0(x_0) e^{\sum_i r_{ii} \tau_i(t)} \prod_{i \neq j} {r_{ij}}^{n_{ij}(t)},
\end{equation}
where $r_{ii} \equiv - \sum_{j(\neq i)} r_{ji}$ is the escape rate, $\tau_i(t) = \int_{\tau=0}^{\tau=t}  \mathrm{d} \tau_i(\tau)$ is the total dwelling time in the state $i$, and $n_{ij}(t) = \int_{\tau=0}^{\tau=t} \mathrm{d} n_{ij}(\tau)$. is the total number of transitions from state $j$ to state $i$.

For the perturbation in the parameter $\lambda \mapsto \lambda + \varepsilon\phi_\lambda(\tau)$, it changes the transition rate as $r_{ij} \mapsto r_{ij}^\varepsilon(\tau) \equiv r_{ij} + \varepsilon \frac{\partial r_{ij}}{\partial\lambda} \phi_\lambda(\tau) + o(\varepsilon)$. Therefore, the trajectory probability of the perturbed system $\mathcal{P}^\varepsilon[X_t]$ is
\begin{equation}
    \mathcal{P}^\varepsilon[X_t] = p_0(x_0) e^{\sum_i r_{ii}^\varepsilon \tau_i(t)} \prod_{i \neq j} \left({r_{ij}^\varepsilon}\right)^{n_{ij}(t)},
\end{equation}
where $r_{ii}^\varepsilon \equiv -\sum_{j(\neq i)} r_{ji}^\varepsilon$. The change on the trajectory probability action is
\begin{subequations}
\begin{align}
    \delta\ln\mathcal{P}[X_t] &\equiv \ln\mathcal{P}^\varepsilon[X_t] - \ln\mathcal{P}[X_t] \\
    &= \varepsilon \int_{\tau=0}^{\tau=t} \phi_\lambda(\tau) \left[ \sum_i \frac{\partial r_{ii}}{\partial \lambda}\mathrm{d}\tau_i(\tau) + \sum_{i \neq j} \frac{\partial\ln r_{ij}}{\partial \lambda} \mathrm{d}n_{ij}(\tau) \right] + o(\varepsilon).
\end{align}
\end{subequations}
Therefore, the conjugate observable is
\begin{subequations}
\begin{align}
    \mathrm{d}\Lambda_\lambda(\tau) &= \dot{\Lambda}_\lambda(\tau)\mathrm{d}\tau \\
    &\equiv \left.\frac{\partial}{\partial \varepsilon}\frac{\delta \ln\mathcal{P}^\varepsilon[X_t]}{\delta \phi_\lambda(\tau)}\right|_{\varepsilon=0} \mathrm{d}\tau \\
    &= \sum_i \frac{\partial r_{ii}}{\partial \lambda}\mathrm{d}\tau_i(\tau) + \sum_{i \neq j} \frac{\partial\ln r_{ij}}{\partial \lambda} \mathrm{d}n_{ij}(\tau) \\
    &= \sum_{i \neq j} \frac{\partial\ln r_{ij}}{\partial\lambda} \mathrm{d}\varepsilon_{ij}(\tau).
\end{align}
\end{subequations}
Then substituting the parameterization $r_{ij} = \exp( b_{ij} + f_{ij}/2 )$, it is straightforward to calculate the conjugate observable for $\{b_{ij}\}$ and $\{f_{ij}\}$:
\begin{subequations}
\begin{align}
    \mathrm{d} \Lambda_{b}(\tau) &= \mathrm{d} \varepsilon_{ij}(\tau) + \mathrm{d} \varepsilon_{ji}(\tau), \\
    \mathrm{d} \Lambda_{f}(\tau) &= \frac{1}{2}\left[ \mathrm{d} \varepsilon_{ij}(\tau) - \mathrm{d} \varepsilon_{ji}(\tau) \right].
\end{align}
\end{subequations}
Due to \cref{SIeq: 2nd moment}, their autocorrelation must be a delta function of the time difference, and the cross-correlations must vanish for different edges. Therefore, the covariance matrix $\Xi = \operatorname{Cov}(\dot{\Lambda}(t_1), \dot{\Lambda}'(t_2))$ must be diagonal
\begin{subequations}
\begin{align}
    \Xi_{b_{ij}}(t_1 - t_2) &= \operatorname{diag} \left\{ \delta(t_1-t_2) a_{ij} \right\}, \\
    \Xi_{f_{ij}}(t_1 - t_2) &= \operatorname{diag} \left\{ \delta(t_1 - t_2) a_{ij}/4 \right\},
\end{align}
\end{subequations}
where we use $r_{ij} \pi_j + r_{ji} \pi_i = a_{ij}$ for steady states.
The Fourier transform of the delta function is a constant function
\begin{equation}
    \int_{-\infty}^{+\infty} e^{\mathrm{i}\omega t} \delta(t) \mathrm{d} t = 1.
\end{equation}
Therefore, the spectral density matrix is $\boldsymbol{\mathcal{L}}(\omega) = \operatorname{diag}\{a_{ij}\}$ or $\operatorname{diag}\{a_{ij}/4\}$ for $\{b_{ij}\}$ or $\{f_{ij}\}$, respectively.

\subsection{Derivation of Eq. (12)} \label{SIsec: derivation Eq13}

To derive Eq. (13) in the main text, we need to calculate the correlation between Poisson noise $\{\mathrm{d} \varepsilon_{ij}(\tau)\}$ and the state-current observable $Q$. For completeness, we review two lemmas from \cite{stutzer2025stochastic}.
\begin{lemma} \label{Lemma: noise-time} (Noise-time Correlations)
    The steady-state correlation between the Poisson noise $\mathrm{d}\varepsilon_{ij}(\tau_1)$ and the dwelling time $\mathrm{d} \tau_k(\tau_2)$ is
    \begin{equation}
        \langle \mathrm{d}\varepsilon_{ij}(\tau_1) \mathrm{d}\tau_k(\tau_2) \rangle = \mathbbold{1}_{\tau_1 < \tau_2} {I_{ij}}^k(\tau_1, \tau_2) r_{ij} \pi_j \mathrm{d}\tau_1 \mathrm{d}\tau_2,
        \label{SIeq: lemma 1}
    \end{equation}
    where ${I_{ij}}^k(\tau_1, \tau_2) = P(k, \tau_2 | i, \tau_1) - P(k, \tau_2 | j, \tau_1)$ depends only on the time difference at steady states.
\end{lemma}
The identificator $\mathbbold{1}_{\tau_1<\tau_2}$ comes from the Markovian nature, where the noise is independent of historical events. The rest part in \cref{SIeq: lemma 1} comes from the joint distribution of $\mathrm{d}\varepsilon_{ij}$ and $\mathrm{d}\tau_k$ given by their definition. We refer \cite{stutzer2025stochastic} for detailed descriptions.

\begin{lemma} \label{Lemma: noise-jump} (Noise-jump Correlations)
    The steady-state correlation between the Poisson noise $\mathrm{d}\varepsilon_{ij}(\tau_1)$ and the stochastic jump $\mathrm{d} n_{kl}(\tau_2)$ is
    \begin{align}
        \langle \mathrm{d}\varepsilon_{ij}(\tau_1) \mathrm{d} n_{kl}(\tau_2) &\rangle = \delta_{ik}\delta_{jl}\delta(\tau_1-\tau_2) r_{ij} \pi_j \mathrm{d}\tau_1 \mathrm{d}\tau_2 \label{SIeq: lemma 2} \\
        &+ \mathbbold{1}_{\tau_1 < \tau_2} {I_{ij}}^l(\tau_1, \tau_2) r_{ij}r_{kl}\pi_j \mathrm{d}\tau_1 \mathrm{d}\tau_2. \nonumber
    \end{align}
\end{lemma}
This lemma is obtained from the linear combination $\langle \mathrm{d}\varepsilon_{ij}(\tau_1) \mathrm{d} n_{kl}(\tau_2) \rangle = \langle \mathrm{d} \varepsilon_{ij}(\tau_1) \mathrm{d} \varepsilon_{kl}(\tau_2) \rangle + r_{kl} \langle \mathrm{d}\varepsilon_{ij}(\tau_1) \mathrm{d}\tau_l(\tau_2) \rangle$.

We first consider state observables $Q(t) = \int_{\tau=0}^{\tau=t} \sum_k g_k \mathrm{d} \tau_k(\tau)$. Using \cref{Lemma: noise-time} and noticing that ${I_{ij}}^k = - {I_{ji}}^k$, we obtain
\begin{align}
    \langle \dot{Q}(\tau_1)  \dot{\Lambda}_{b_{ij}}(\tau_2) \rangle &= \frac{\langle \mathrm{d} Q(\tau_1) (\mathrm{d} \varepsilon_{ij}(\tau_2) + \mathrm{d} \varepsilon_{ji}(\tau_2)) \rangle}{\mathrm{d}\tau_1 \mathrm{d}\tau_2} \nonumber \\
    &= \mathbbold{1}_{\tau_1>\tau_2} \left( r_{ij}\pi_j \sum_k g_k {I_{ij}}^k + r_{ji}\pi_i \sum_k g_k {I_{ji}}^k \right) \nonumber \\
    &= \mathbbold{1}_{\tau_1>\tau_2} \left( r_{ij}\pi_j \sum_k g_k {I_{ij}}^k - r_{ji}\pi_i \sum_k g_k {I_{ij}}^k \right) \nonumber \\
    &= j_{ij} \mathbbold{1}_{\tau_1>\tau_2} \sum_k g_k {I_{ij}}^k(\tau_1, \tau_2) \nonumber \\
    &= j_{ij} G(\tau_1, \tau_2),
\end{align}
where $G(\tau_1, \tau_2) \equiv \mathbbold{1}_{\tau_1>\tau_2} \sum_k g_k {I_{ij}}^k(\tau_1, \tau_2)$. Similarly, we have $\langle \dot{Q}(\tau_1) \dot{\Lambda}_{f_{ij}}(\tau_2) \rangle = \frac{1}{2} a_{ij} G(\tau_1, \tau_2)$.
Therefore, the time-domain response function satisfies $R_{b_{ij}}/R_{f_{ij}} = 2j_{ij}/a_{ij}$. Noticing that only the term $G(\tau_1, \tau_2)$ is time-dependent. As a consequence, $\mathcal{G}(\omega) = \int_{-\infty}^{+\infty} e^{\mathrm{i}\omega \tilde{t}} G(\tilde{t}) \mathrm{d} \tilde{t}$ also vanishes after the Fourier transform. Therefore, we have $\mathcal{R}_{b_{ij}}/\mathcal{R}_{f_{ij}} = 2j_{ij}/a_{ij}$ for state observables.

Next, we consider current observables $Q(t) = \int_{\tau=0}^{\tau=t} \sum_{k \neq l} h_{kl} \mathrm{d} n_{kl}(\tau)$ with antisymmetric coefficients $h_{kl} = - h_{lk}$. Using \cref{Lemma: noise-jump}, ${I_{ij}}^k = - {I_{ji}}^k$, and $h_{kl} = -h_{lk}$, we obtain the following relation:
\begin{align}
    \langle \dot{Q}(\tau_1) \dot{\Lambda}_{b_{ij}}(\tau_2) \rangle
    &= \frac{\langle \mathrm{d} Q(\tau_1) (\mathrm{d} \varepsilon_{ij}(\tau_2) + \mathrm{d} \varepsilon_{ji}(\tau_2)) \rangle}{\mathrm{d}\tau_1 \mathrm{d}\tau_2} \nonumber \\
    &= \mathbbold{1}_{\tau_1=\tau_2} \left(h_{ij}r_{ij}\pi_j + h_{ji}r_{ji}\pi_i \right) + \mathbbold{1}_{\tau_1>\tau_2} \left( r_{ij}\pi_j \sum_{k \neq l} h_{kl}r_{kl}{I_{ij}}^l + r_{ji}\pi_i \sum_{k \neq l} h_{kl}r_{kl}{I_{ji}}^l \right) \nonumber \\
    &= \mathbbold{1}_{\tau_1=\tau_2} j_{ij} h_{ij} + \mathbbold{1}_{\tau_1>\tau_2} j_{ij} \sum_{k\neq l} h_{kl}r_{kl}{I_{ij}}^l \nonumber \\
    &= j_{ij} H(\tau_1, \tau_2),
\end{align}
where we define $H(\tau_1, \tau_2) \equiv ( \mathbbold{1}_{\tau_1=\tau_2} h_{ij} + \mathbbold{1}_{\tau_1>\tau_2} \sum_{k\neq l} h_{kl}r_{kl}{I_{ij}}^l )$ for convenience. Similarly, $\langle \dot{Q}(\tau_1) \dot{\Lambda}_{f_{ij}}(\tau_2) \rangle = \frac{1}{2}a_{ij} H(\tau_1, \tau_2)$ holds for entropic force perturbations. Therefore, the time-domain response function satisfies $R_{b_{ij}}/R_{f_{ij}} = 2j_{ij}/a_{ij}$. Noticing that only the term $H(\tau_1, \tau_2)$ is time-dependent. As a consequence, $\mathcal{H}(\omega) = \int_{-\infty}^{+\infty} e^{\mathrm{i}\omega \tilde{t}} H(\tilde{t}) \mathrm{d} \tilde{t}$ also vanishes after the Fourier transform. Therefore, we have $\mathcal{R}_{b_{ij}}/\mathcal{R}_{f_{ij}} = 2j_{ij}/a_{ij}$ for the current observables.

Notice that similar relations are also proved in \cite{kwon2025fluctuation} using eigenvalue analysis. We would like to emphasize that the response-ratio relation proved in \cite{kwon2025fluctuation} is the time-integral version of our relation. So, our results offer more detailed information about the response function and make it possible to extend time-domain theories into the frequency-domain.

\section{Derivations for continuous systems} \label{SIsec: inequality Langevin}

\subsection{Conjugate observable for force perturbations} \label{SIsec: Lambda_F}

We start with the definition of the response function. For an observable $\langle \dot{Q} \rangle$, the steady-state response to perturbations in the parameter $\lambda(x) \mapsto \lambda(x) + \varepsilon\phi_\lambda(x, t)$ is given by
\begin{equation}
    \delta\langle \dot{Q}(t) \rangle = \varepsilon \int_{t=0}^{t=\tau} R_\lambda(x, t-\tau) \phi_\lambda(x, \tau) \mathrm{d} \tau + o(\varepsilon).
\end{equation}
The response function is defined as $R_\lambda(x, t-\tau) \equiv \frac{\delta\langle \dot{Q}(t) \rangle}{\delta \phi_\lambda(x, \tau)}$. It represents the response at time $t$ stimulated by the perturbation at time $\tau$. It is a function of $t-\tau$ due to the time-translation invariance in steady states. The response function can be written as the correlation between the observable and a conjugate observable
\begin{equation}
    R_\lambda(x, t - \tau) = \operatorname{Cov}(\dot{Q}(t), \dot{\Lambda}_\lambda(x, \tau)),
\end{equation}
where $\dot{\Lambda}_\lambda(x, \tau) \equiv \left.\frac{\partial}{\partial\varepsilon}\frac{\delta\ln\mathcal{P}^\varepsilon[X_t]}{\delta\phi_\lambda(x, \tau)}\right|_{\varepsilon=0}$.

For the overdamped Langevin dynamics, the trajectory probability is
\begin{equation}
    \mathcal{P}[X_t] = p_0(x_0) \mathcal{N} e^{-\mathcal{A}[X_t]},
\end{equation}
where $p_0(x_0)$ is the initial probability distribution, $\mathcal{N}$ is the normalizer, and $\mathcal{A}[X_t]$ is the path action of the trajectory $X_t$. The action $\mathcal{A}[X_t]$ takes the following form:
\begin{equation}
    \mathcal{A}[X_t] = \int_{\tau=0}^{\tau=t} \frac{[\dot{x}_\tau - v(x_\tau)]^2}{4 \mu(x_\tau) T(x_\tau)} \mathrm{d}\tau,
\end{equation}
where $v(x) \equiv \mu(x) F(x) + \partial_x [\mu(x)T(x)]$ is the effective drift velocity. The second term in $v(x)$ represents the spurious drift that arises from the conversion from the anti-Ito product to the Ito product. The normalizer is given by $\mathcal{N} = \int \mathcal{D}[X_t] e^{-\mathcal{A}[X_t]}$. The path probability of the perturbed system is given by $\mathcal{P}^\varepsilon[X_t] = p_0(x_0) \mathcal{N}^\varepsilon e^{-\mathcal{A}^\varepsilon[X_t]}$.

We first consider the perturbation on the force $F(x) \mapsto F^\varepsilon(x) \equiv F(x) + \varepsilon \phi_F(x, \tau)$. In this case, the conjugate observable is
\begin{subequations}
\begin{align}
    \dot{\Lambda}_F(x, \tau) &\equiv \left.\frac{\partial}{\partial\varepsilon}\frac{\delta\ln\mathcal{P}^\varepsilon[X_t]}{\delta\phi_F(x, \tau)}\right|_{\varepsilon=0} \\
    &= \left\langle \left. \frac{\partial}{\partial\varepsilon}\frac{\delta\mathcal{A}^\varepsilon[X_t]}{\delta\phi_F(x, \tau)} \right|_{\varepsilon=0} \right\rangle - \left. \frac{\partial}{\partial\varepsilon}\frac{\delta\mathcal{A}^\varepsilon[X_t]}{\delta\phi_F(x, \tau)} \right|_{\varepsilon=0},
\end{align}
\end{subequations}
where we use $\frac{\partial}{\partial\varepsilon}\frac{\delta \mathcal{N}^\varepsilon}{\delta \phi_F(x, \tau)} = \left\langle \frac{\partial}{\partial\varepsilon}\frac{\delta\mathcal{A}^\varepsilon[X_t]}{\delta\phi_F(x, \tau)} \right\rangle$. The perturbed action is
\begin{equation}
    \mathcal{A}^\varepsilon[X_t] = \int_{\tau=0}^{\tau=t} \frac{[\dot{x}_\tau - v^\varepsilon(x_\tau, \tau)]^2}{4 \mu(x_\tau) T(x_\tau)} \mathrm{d}\tau,
\end{equation}
where $v^\varepsilon(x, \tau) = v(x) + \varepsilon\mu(x)\phi_F(x, \tau)$. It gives the function derivative of the action:
\begin{subequations}
\begin{align}
    \left. \frac{\partial}{\partial\varepsilon}\frac{\delta\mathcal{A}^\varepsilon[X_t]}{\delta\phi_F(x, \tau)} \right|_{\varepsilon=0} &= -\left. \frac{\partial}{\partial\varepsilon} \left[\varepsilon \frac{\dot{x}_\tau - v^\varepsilon(x_\tau)}{2T(x_\tau)} \delta(x-x_\tau) \right] \right|_{\varepsilon=0} \\
    &= -\frac{\dot{x_\tau} - v(x_\tau)}{2 T(x_\tau)} \delta(x-x_\tau) \\
    &= -\sqrt{\frac{\mu(x_\tau)}{2T(x_\tau)}} \xi_\tau \delta(x-x_\tau),
\end{align}
\end{subequations}
where we use the Langevin equation in the third equality. This implies that the average always vanishes:
\begin{equation}
    \left\langle \left. \frac{\partial}{\partial\varepsilon}\frac{\delta\mathcal{A}^\varepsilon[X_t]}{\delta\phi_F(x, \tau)} \right|_{\varepsilon=0} \right\rangle = 0,
\end{equation}
since $\langle \xi(x_\tau) \rangle = 0$. Therefore, the conjugate observable is
\begin{equation}
    \dot{\Lambda}_F(x, \tau) = \sqrt{\frac{\mu(x_\tau)}{2T(x_\tau)}} \xi_\tau \delta(x-x_\tau).
\end{equation}

Notice that the covariance function of $\dot{\Lambda}_F$ contains delta functions on space and time,
\begin{subequations}
\begin{align}
    \operatorname{Cov}(\dot{\Lambda}_F(x_1, \tau_1), \dot{\Lambda}_F(x_2, \tau_2)) &= \sqrt{\frac{\mu(x_1)\mu(x_2)}{4 T(x_1) T_(x_2)}} \langle \xi_{\tau_1} \xi_{\tau_2} \delta(x-x_1) \delta(x-x_2) \rangle \\
    &= \frac{\mu(x_1)}{2T(x_1)} \pi(x_1) \delta(\tau_1-\tau_2)\delta(x_1-x_2),
\end{align}
\end{subequations}
where we use $\langle \xi_{\tau_1} \xi_{\tau_2} \rangle = \delta(\tau_1-\tau_2)$ and $\langle \delta(x-x_1) \rangle = \pi(x_1)$, and $\pi$ is the steady-state distribution.

\subsection{Inequality for infinite dimensional Schur complement} \label{SIsec: inf Schur inequality}

For strictness and completeness, we give a statement for the infinite dimensional Schur complement inequality, which we use to derive the FRIs for Langevin systems. One may roughly consider the continuous case as infinite dimensional matrices, then the results follow straightforwardly from the discrete case.

Consider the infinite dimensional vector $\boldsymbol{V}(t) = (\dot{Q}(t), \{\dot{\Lambda}_\lambda(x, t)\}_x)$. The Fourier transform of its steady-state covariance matrix constitutes an operator (kernel function) that acts on the function pair $(a, b(x)) \in \mathbb{C} \oplus L^2$ (where $\mathbb{C}$ represents the complex field and $L^2$ represents square integrable functions):
\begin{equation}
    \mathbb{S}(\omega) = \begin{pmatrix}
        \mathcal{S}(\omega) & \mathcal{B}(\omega) \\
        \mathcal{B}^\dagger(\omega) & \mathcal{L}(\omega)
    \end{pmatrix}.
\end{equation}
$\mathcal{S}(\omega) = \int e^{\mathrm{i}\omega t} \operatorname{Cov}(\dot{Q}(t), \dot{Q}(0)) \mathrm{d}t$ is a real number due to time-translation symmetry. $\mathcal{B}(\omega)$ is a functional that maps a $L^2$ function to a complex number: $(\mathcal{B}b)(\omega) = \int \mathcal{R}_\lambda(x, \omega)b(x) \mathrm{d}x$, where $\mathcal{R}_\lambda$ is the Fourier transform of the response function $\mathcal{R}_\lambda(x, \omega) = \int e^{\mathrm{i}\omega \tilde{t}} R_\lambda(x, \tilde{t}) \mathrm{d}\tilde{t}$. $\mathcal{R}_\lambda(x, \omega)$ is the kernel of the functional $\mathcal{B}(\omega)$. $\mathcal{B}^\dagger$ is the adjoint operator of $\mathcal{B}$: for any $a \in \mathbb{C}$, $(\mathcal{B}^\dagger a)(x) = a^* \mathcal{R}_\lambda(x, \omega)$, s.t.,
\begin{equation}
    (\mathcal{B}b)^* a = \int b^*(x) (\mathcal{B}^\dagger a)(x) \mathrm{d}x.
\end{equation}
$\mathcal{L}(\omega)$ is the spectrum density operator of $\Lambda_\lambda(\dot{x}, t)$. The kernel of the operator $\mathcal{L}(\omega)$ is
\begin{equation}
    \mathcal{L}(x, x', \omega) = \int e^{\mathrm{i}\omega t} \operatorname{Cov}(\dot{\Lambda}_\lambda(x, t), \dot{\Lambda}_\lambda(x', 0)) \mathrm{d}t.
\end{equation}
Notice that the covariance contains space-time delta functions. So, the operator $\mathcal{L}(\omega) = \mathcal{L}_\lambda(x)\delta(x-x')$ is also independent of $\omega$.

For any function pair $(a, b(x)) \in \mathbb{C} \oplus L^2$, the steady-state covariance matrix of $\boldsymbol{V}(t)$ must be positive semi-definite, that is,
\begin{equation}
    a^* \mathcal{S}(\omega) a + a^* (\mathcal{B}b) + (\mathcal{B}b) a + \int |b(x)|^2 \mathcal{L}_\lambda(x) \mathrm{d}x \ge 0,
\end{equation}
where $a^*$ is the complex conjugate of $a$.

The schur complement for Hilbert space gives the inequality
\begin{equation}
    \mathcal{S}(\omega) - \mathcal{B} \mathcal{L}_\lambda^{-1} \mathcal{B}^\dagger \ge 0.
\end{equation}
It is an operator inequality in the sense that the operator $\mathcal{S}(\omega) - \mathcal{B} \mathcal{L}_\lambda^{-1} \mathcal{B}^\dagger$ is positive semi-definite, i.e., $a^*(\mathcal{S}(\omega) - \mathcal{B} \mathcal{L}_\lambda^{-1} \mathcal{B}^\dagger)a \ge 0$ for any $a \in \mathbb{C}$. The operator inequality gives the following real number inequality
\begin{equation}
    \mathcal{S}(\omega) - \int \frac{|\mathcal{R}_\lambda(x, \omega)|^2}{\mathcal{L}_\lambda(x)} \mathrm{d}x \ge 0
\end{equation}
A direct rearrangement yields the inequality Eq. (A10a) in the End Matter:
\begin{equation}
    \int \frac{|\mathcal{R}_\lambda(x, \omega)|^2}{\mathcal{L}_\lambda(x)} \mathrm{d}x \le \mathcal{S}(\omega).
\end{equation}

\subsection{\texorpdfstring{Specific forms of $\mathcal{L}_\lambda(x)$}{Specific forms of L(x)}}

For perturbations on the parameter $F(x)$, we have
\begin{equation}
    \operatorname{Cov}(\dot{\Lambda}_F(x_1, \tau_1), \dot{\Lambda}_F(x_2, \tau_2)) = \frac{\mu(x_1)\pi(x_1)}{2T(x_1)}  \delta(\tau_1-\tau_2)\delta(x_1-x_2).
\end{equation}
Its Fourier transform is
\begin{subequations}
\begin{align}
    \mathcal{L}(x, x', \omega) &= \int e^{\mathrm{i}\omega t} \operatorname{Cov}(\dot{\Lambda}_F(x, t), \dot{\Lambda}_F(x', 0)) \mathrm{d}t \\
    &= \frac{\mu(x)\pi(x)}{2 T(x)} \delta(x - x').
\end{align}
\end{subequations}

Therefore, the FRI for local force perturbations becomes the following:
\begin{equation}
    \int \frac{|\mathcal{R}_F(x, \omega)|^2}{\mathcal{L}_F(x)} \mathrm{d}x = \int \frac{|\mathcal{R}_F(x, \omega)|^2}{\mu(x)\pi(x) / (2T(x))} \mathrm{d}x \le \mathcal{S}(\omega). \label{SIeq: FRIs F}
\end{equation}
It gives the relation $\mathcal{L}_F = \pi\mu / (2T)$.

Now we derive finite-frequency FRIs for perturbations in $\ln\mu(x)$ and $T(x)$. Similarly to the discrete case, we utilize the ratio of frequency-domain response functions to derive the FRIs for these two parameters. This ratio has been derived recently in \cite{chun2026fluctuation}, we restate the results for completeness and extend it into the frequency domain.

Consider the time dependent perturbations $\lambda(x) \mapsto \lambda^\varepsilon(x, \tau) \equiv \lambda(x) + \varepsilon\phi(x, \tau)$. For the state-current observable
\begin{equation}
    Q(x, t) = \int_{\tau=0}^{\tau=t} \left[u(x_\tau) + \dot{x}_\tau \circ w(x_\tau)\right] \mathrm{d}\tau,
\end{equation}
the recent research \cite{chun2026fluctuation} found that the ratio of the steady-state response function of state-current observables are independent of both the perturbation and the observable:
\begin{equation}
    \frac{R_F(x, t-\tau)}{R_\lambda(x, t-\tau)} = \frac{\mu(x) \pi(x)}{N_\lambda(x)}, \label{SIeq: response ratio}
\end{equation}
where $\lambda$ and $N_\lambda(x)$ can be read from \cref{SItab: response ratio}. Since the right-hand-side of \cref{SIeq: response ratio} is independent of $(t-\tau)$, the arrangement and Fourier transform lead to the ratio relation for frequency-domain relation:
\begin{equation}
    \frac{\mathcal{R}_F(x, \omega)}{\mathcal{R}_\lambda(x, \omega)} = \frac{\mu(x) \pi(x)}{N_\lambda(x)}. \label{SIeq: freq response ratio}
\end{equation}

\begin{table}[htbp]
    \centering
    \begin{tabular}{cc}
        \toprule
        $\lambda(x)$ & $N_\lambda(x)$ \\
        \midrule
        $\ln\mu(x)$ & $j_\text{ss}(x)$ \\
        $F(x)$ & $\pi(x)\mu(x)$ \\
        $T(x)$ & $-\mu(x)\partial_x\pi(x)$ \\
        \bottomrule
    \end{tabular}
    \caption{Perturbed parameter $\lambda(x)$ and its corresponding $N_\lambda(x)$}
    \label{SItab: response ratio}
\end{table}

Substituting \cref{SIeq: freq response ratio} into \cref{SIeq: FRIs F} leads to $\mathcal{L}_{\ln\mu} = {j_\text{ss}}^2 / (2\pi \mu T)$ and $\mathcal{L}_T = \pi\mu(\partial_x \ln\pi)^2 / (2T)$.

\section{Simulations} \label{SIsec: simulation}

In the main text, we use numerical simulations and exact numerical linear algebra to verify the finite-frequency response relations. All Markov jump processes are simulated from their unperturbed steady states. We use the column convention for the generator, $\dot{\bm p}=R\bm p$, where $R_{ij}$ is the transition rate from state $j$ to state $i$. For each undirected edge $e$, we choose a positive orientation from the tail state $s_e^-$ to the head state $s_e^+$ and write
\begin{equation}
    r_e^+ = \exp\left(b_e + \frac{f_e}{2}\right),
    \qquad
    r_e^- = \exp\left(b_e - \frac{f_e}{2}\right).
\end{equation}
The steady-state current and traffic on edge $e$ are
\begin{equation}
    j_e = r_e^+ \pi_{s_e^-} - r_e^- \pi_{s_e^+},
    \qquad
    a_e = r_e^+ \pi_{s_e^-} + r_e^- \pi_{s_e^+}.
\end{equation}
The entropy production rate and pseudo-entropy-production rate are computed as
\begin{equation}
    \dot{\sigma} = \sum_e j_e \ln \frac{r_e^+ \pi_{s_e^-}}{r_e^- \pi_{s_e^+}},
    \qquad
    \dot{\sigma}_{\mathrm{pseudo}} = \sum_e \frac{2j_e^2}{a_e}.
\end{equation}

For the trajectory-based calculations, we use a discrete-time binning of the continuous-time dynamics. In each time interval $[m\Delta t,(m+1)\Delta t]$, at most one jump is sampled. If the system is in state $s$, the jump probability in this bin is approximated by $r_{\mathrm{esc}}(s)\Delta t$, where $r_{\mathrm{esc}}(s)$ is the escape rate from state $s$. Conditional on a jump, the outgoing transition is sampled with probability proportional to its transition rate. The initial state of each trajectory is sampled from the steady-state distribution $\bm{\pi}$. In all simulations, we check that $\max_s r_{\mathrm{esc}}(s)\Delta t$ is small enough for the one-jump-per-bin approximation.

For a trajectory observable $Q$, we discretize its rate as
\begin{equation}
    q_m^{(\alpha)} \equiv \frac{\Delta Q_m^{(\alpha)}}{\Delta t},
\end{equation}
where $\Delta Q_m^{(\alpha)}$ is the increment of $Q$ along trajectory $\alpha$ during the interval $[m\Delta t,(m+1)\Delta t]$. The fluctuation around the steady-state mean is defined as
\begin{equation}
    \delta q_m^{(\alpha)} \equiv q_m^{(\alpha)}-\bar q,
    \qquad
    \bar q \equiv \frac{1}{N_{\mathrm{traj}}N_t} \sum_{\alpha=1}^{N_{\mathrm{traj}}} \sum_{m=0}^{N_t-1} q_m^{(\alpha)}.
\end{equation}
Here $N_t$ is the number of time bins, $T_{\mathrm{obs}}=N_t\Delta t$ is the trajectory duration, and $N_{\mathrm{traj}}$ is the number of independently sampled trajectories.

The spectral density of $\dot Q$ is estimated by the finite-time periodogram
\begin{equation}
    \mathcal{S}_Q(\omega_j) = \frac{\Delta t}{N_t N_{\mathrm{traj}}} \sum_{\alpha=1}^{N_{\mathrm{traj}}} \left| \sum_{m=0}^{N_t-1} \delta q_m^{(\alpha)} e^{\mathrm{i}\omega_j m\Delta t} \right|^2,
\end{equation}
where
\begin{equation}
    \omega_j=\frac{2\pi j}{N_t\Delta t}
\end{equation}
are the discrete Fourier frequencies. This estimator is a direct trajectory-sampling implementation of the definition of the power spectral density and does not require prior knowledge of the transition-rate matrix beyond the simulated dynamics.

The response functions are estimated from the conjugate observables associated with the path-probability score. For an edge-wise barrier perturbation $b_e\mapsto b_e+\epsilon\phi(t)$, the infinitesimal conjugate observable is
\begin{equation}
    \mathrm{d}\Lambda_{b_e}(t) = \mathrm{d} n_e^+(t) + \mathrm{d} n_e^-(t) - r_e^+ \mathrm{d} \tau_{s_e^-}(t) - r_e^- \mathrm{d} \tau_{s_e^+}(t).
\end{equation}
For an edge-wise entropic-force perturbation $f_e\mapsto f_e+\epsilon\phi(t)$, it is
\begin{equation}
\mathrm{d}\Lambda_{f_e}(t) = \frac{1}{2}\mathrm{d} n_e^+(t) - \frac{1}{2}\mathrm{d} n_e^-(t) - \frac{1}{2}r_e^+ \mathrm{d}\tau_{s_e^-}(t) + \frac{1}{2}r_e^- \mathrm{d}\tau_{s_e^+}(t).
\end{equation}
In the simulations, we discretize $\dot{\Lambda}_{\lambda}(t)$ in the same way as $\dot Q(t)$ and compute the causal covariance
\begin{equation}
C_{Q,\lambda}(\ell\Delta t)
=
\frac{1}{N_{\mathrm{traj}}(N_t-\ell)}
\sum_{\alpha=1}^{N_{\mathrm{traj}}}
\sum_{m=0}^{N_t-\ell-1}
\delta q_{m+\ell}^{(\alpha)}
\delta \lambda_m^{(\alpha)},
\qquad
\ell\geq 0,
\end{equation}
where $\delta \lambda_m^{(\alpha)}$ denotes the discretized fluctuation of $\dot{\Lambda}_{\lambda}$ in the $m$-th bin. The finite-frequency response is then computed as
\begin{equation}
    \mathcal{R}_{\lambda}(\omega_j) = \Delta t \sum_{\ell=0}^{N_t-1} C_{Q,\lambda}(\ell\Delta t) e^{\mathrm{i}\omega_j \ell\Delta t}.
\end{equation}
In the numerical implementation, the above causal correlations and Fourier transforms are evaluated by fast Fourier transforms.

For Fig. 1 in the main text, we sample random complete-graph Markov jump processes to verify Eqs. (11a) and (11b). We consider state-space sizes $N = 3, 4, 5, 6$. For each $N$, we sample $8$ independent rate matrices. For every undirected edge $e$, the kinetic parameter and force parameter are sampled as
\begin{equation}
    b_e \sim \mathrm{Unif}[-0.65,0.55] + \ln(0.45),
    \qquad
    f_e \sim \mathrm{Unif}[-2,2],
\end{equation}
and the rates are constructed from $r_e^\pm = \exp(b_e\pm f_e/2)$. For each rate matrix, we sample $8$ generic trajectory observables and $8$ state-current observables. A generic trajectory observable has the form
\begin{equation}
    Q(t) = \int_0^t g_{x_\tau}\mathrm{d}\tau + \sum_e \left[ h_e^+ n_e^+ (t) + h_e^- n_e^- (t) \right],
\end{equation}
where $g_i$, $h_e^+$, and $h_e^-$ are independently sampled random coefficients. For the state-current observables, the jump coefficients are constrained by $h_e^- = -h_e^+$. The coefficients $g_i$ are normalized to have zero mean and unit variance for each observable, while the jump coefficients are sampled from a normal distribution with standard deviation $0.75$. The simulations use
\begin{equation}
    \Delta t=0.01,
    \qquad
    N_t=1024,
    \qquad
    T_{\mathrm{obs}}=10.24,
    \qquad
    N_{\mathrm{traj}}=5000.
\end{equation}
For each observable, we compute
\begin{equation}
    \frac{1}{\mathcal{S}_Q(\omega)}
    \sum_e
    \frac{|\mathcal{R}_{b_e}(\omega)|^2}{a_e},
    \qquad
    \frac{1}{\mathcal{S}_Q(\omega)}
    \sum_e
    \frac{4|\mathcal{R}_{f_e}(\omega)|^2}{a_e},
\end{equation}
and plot these quantities as scatter plots over different models, observables, and frequencies.

For Fig. 2, we simulate a channel-resolved spinful quantum dot coupled to two electronic reservoirs. The four dot states are $0$, $\uparrow$, $\downarrow$, and $2$, with energies
\begin{equation}
    E_0 = 0,
    \qquad
    E_\uparrow = \epsilon-\frac{\Delta}{2},
    \qquad
    E_\downarrow = \epsilon+\frac{\Delta}{2},
    \qquad
    E_2 = 2\epsilon+U.
\end{equation}
The Fermi function of lead $\alpha = L,R$ is
\begin{equation}
    f_\alpha(E) = \frac{1}{1+\exp[\beta(E-\mu_\alpha)]}.
\end{equation}
For each lead-resolved channel, the positive direction is chosen as electron tunneling from the lead into the dot. The corresponding channel rates are
\begin{subequations}
\begin{align}
    r^\alpha_{\sigma,0} &= \Gamma_{\alpha\sigma}^{(0)} f_\alpha(E_\sigma-E_0),
    \\
    r^\alpha_{0,\sigma} &= \Gamma_{\alpha\sigma}^{(0)} \left[ 1-f_\alpha(E_\sigma-E_0) \right],
    \\
    r^\alpha_{2,\sigma} &= \Gamma_{\alpha\bar\sigma}^{(1)} f_\alpha(E_2-E_\sigma),
    \\
    r^\alpha_{\sigma,2} &= \Gamma_{\alpha\bar\sigma}^{(1)} \left[ 1-f_\alpha(E_2-E_\sigma) \right].
\end{align}
\end{subequations}
The parameters are
\begin{equation}
    \beta = 1,
    \qquad
    \epsilon = 0,
    \qquad
    U = 2.0,
    \qquad
    \Delta_Z = 0.6,
    \qquad
    \mu_L = 2.5,
    \qquad
    \mu_R = -1.5,
\end{equation}
and
\begin{equation}
    \left(
    \Gamma_{L\uparrow}^{(0)},
    \Gamma_{L\downarrow}^{(0)},
    \Gamma_{R\uparrow}^{(0)},
    \Gamma_{R\downarrow}^{(0)},
    \Gamma_{L\uparrow}^{(1)},
    \Gamma_{L\downarrow}^{(1)},
    \Gamma_{R\uparrow}^{(1)},
    \Gamma_{R\downarrow}^{(1)}
    \right) = \left(
    1.0, 0.7, 0.6, 1.1, 0.8, 1.2, 1.0, 0.75
    \right).
\end{equation}
The measured observable is the net particle current into the right lead. Therefore, for a right-lead channel, a positive jump has an increment of $-1$, while a negative jump has an increment of $+1$. The perturbed edge is the right-lead channel connecting $0$ and $\downarrow$. We estimate both $\mathcal{R}*{b_e}(\omega)$ and $\mathcal{R}*{f_e}(\omega)$ from the same set of unperturbed trajectories and compare their ratio with
\begin{equation}
    \frac{\mathcal{R}_{b_e}(\omega)}{\mathcal{R}_{f_e}(\omega)} = \frac{2j_e}{a_e} = 2\tanh\frac{A_e}{2}.
\end{equation}
The simulations use
\begin{equation}
    \Delta t = 0.02,
    \qquad
    N_t = 4096,
    \qquad
    T_{\mathrm{obs}} = 81.92,
\end{equation}
with $N_{\mathrm{traj}} = 5000$ independent trajectories for the data shown in the main text.

\bibliography{supplement}